\documentclass[a4paper,12pt]{article}
\usepackage{epsfig}

\def\Title#1{\begin{center} {\Large #1 } \end{center}}
\def\Author#1{\begin{center}{ \sc #1} \end{center}}
\def\Address#1{\begin{center}{ \it #1} \end{center}}

\newenvironment{Abstract}{\begin{quotation}  }{\end{quotation}}
\newenvironment{Presented}{\begin{quotation} \begin{center} 
             PRESENTED AT\end{center}\bigskip 
      \begin{center}\begin{large}}{\end{large}\end{center} \end{quotation}}

\def\beq{\begin{equation}}
\def\eeq#1{\label{#1}\end{equation}}
\def\eeqn{\end{equation}}

\def\beqa{\begin{eqnarray}}
\def\eeqa#1{\label{#1}\end{eqnarray}}
\def\eeqan{\end{eqnarray}}

\let\bar=\overbar

\def\Dslash{\not{\hbox{\kern-4pt $D$}}}
\def\dslash{\not{\hbox{\kern-2pt $\del$}}}

\def\msb{{\bar{\ssstyle M \kern -1pt S}}}

\begin{document}

\begin{titlepage}

\vfill
\Title{Hadronic interactions at ultra high energies \\-- tests with the Pierre Auger Observatory}
\vfill
\Author{Sofia Andringa$^a$ \index{Andringa, S.}, for the Pierre Auger Collaboration$^b$}
\Address{a) LIP - Laborat\'orio de F\'{\i}sica Experimental de Part\'{\i}culas \\
Av. Prof. Gama Pinto, 2 (3I's), 1649-003 Lisboa, PORTUGAL\\
b) Observatory Pierre Auger, Av. San Mart{\'\i}n Norte 304, 5613 Malarg\"ue, Argentina\\
Full author list: $http://www.auger.org/archive/authors\_2017\_06.html$}
\vfill
\begin{Abstract}
The Pierre Auger Observatory is a hybrid detector for cosmic rays with $E>1$\,EeV. From the gathered data we estimated the proton-proton cross-section at $\sqrt{s}=55$~TeV and tested other features of the hadronic interaction models, which use extrapolations from the LHC energy. The electromagnetic component, carrying most of the energy of the shower, is precisely measured using fluorescence telescopes, while the hadronic backbone of the shower is indirectly tested by measuring the muons arriving to the surface detector. The analyses show that models fail to describe these two components consistently, predicting too few muons at the ground.
\end{Abstract}
\vfill
\begin{Presented}
Presented at EDS Blois 2017, Prague, \\ Czech Republic, June 26-30, 2017
\end{Presented}
\vfill
\end{titlepage}
\def\thefootnote{\fnsymbol{footnote}}
\setcounter{footnote}{0}





%


\section{Electromagnetic shower observed by FD}

Each of the four sites of the Fluorescence Detector (FD) of the Pierre Auger Observatory covers 180$^\circ$ in azimuth and $[2^\circ,32^\circ]$ in elevation, imaging the shower development over the Surface Detector (SD) array. When the depth of shower maximum ($X_\mathrm{max}$) is in the field of view, integration of a Gaisser-Hillas distribution fit to the observed shower profile, yields a calorimetric energy measurement. To increase the acceptance for lower energies, three extra telescopes are pointed to higher elevations. 

\begin{figure}[htb]
\begin{center}
\begin{minipage}[b]{0.65\linewidth}
\epsfig{file=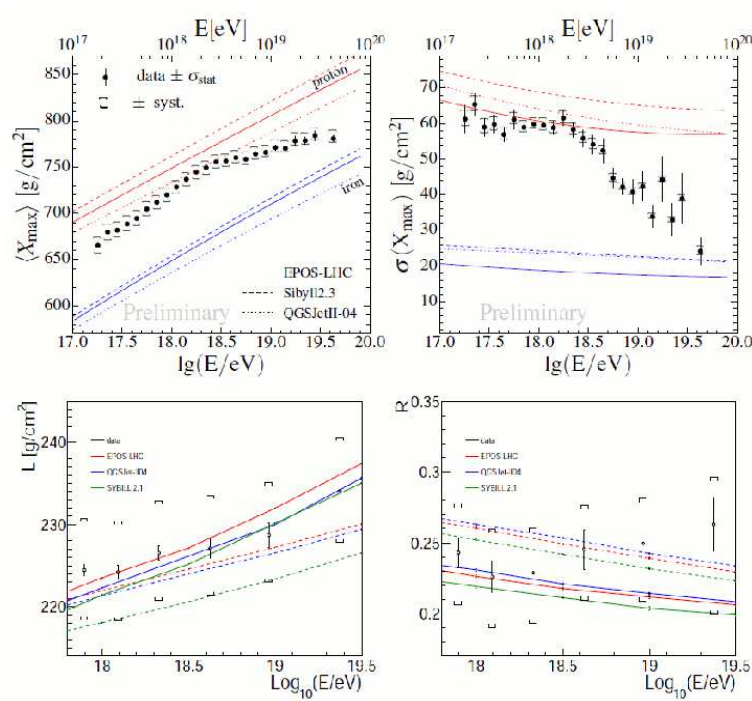,width=1.00\linewidth}
\end{minipage}
\hspace{-0.25cm}\hfill
\begin{minipage}[b]{0.35\linewidth}
\caption{Energy evolution of FD variables compared to models: the mean and $\sigma$ of $X_\mathrm{max}$ (from \cite{xmax}, top); the profile shape variables (from \cite{Francisco}, bottom).} 
\label{fig:FD}
\vspace{0.5cm}
\end{minipage}
\end{center}
\end{figure}

$X_\mathrm{max}$ distributions have been measured down to  $\log_{10}(E/\mathrm{eV})\sim17.2$ (close to the LHC $\sqrt{s}$), in non-biased samples~\cite{xmax}. The evolution of their mean and variance (Fig. \ref{fig:FD}top) indicates a proton dominated beam at $\log_{10}(E/\mathrm{eV})\sim18.25$, with mixed heavy (light) composition at lower (higher) energies. 
Fitting templates for different primary nuclei, obtained with each hadronic interaction model 
gives common trends. However, the width of the distributions is better described in EPOS-LHC~\cite{EPOS}, while in QGSJet-II.04~\cite{QGS} the shower-to-shower fluctuations are too small, or an evolution with primary mass is not fast enough.

It is instructive to decompose $X_\mathrm{max}=X_1+\Delta$. The first interaction depth, $X_1$, gives rise to an exponential tail, corresponding to the lowest primary interaction cross-section. The proton-proton inelastic cross-section has been measured and is in good agreement with the  extrapolations from the LHC data, up to $\sqrt{s}=55.5\pm3.6$~TeV~\cite{xsec}. The development depth, $\Delta$, is reflected in the shape of each shower. To assess a very precise shape~\cite{Francisco}, each shower is translated and normalized relative to its maximum and an average profile is created for each energy bin, compatible within 1\% with a Gaisser-Hillas funtion for $X-X_\mathrm{max}\in[-300,+200]~\mathrm{g/cm^2}$. Fig. \ref{fig:FD}(bottom) shows the resulting shape parameters: $L$, the main scale for the calorimetric energy, follows the expectations; $R$, more sensitive to the shower start, has a faster energy evolution.


\section{The muonic shower component detected by SD}

The SD is composed of 1660 water Cherenkov detectors, sampling the particles arriving at the ground over an area of 3000~km$^2$, with higher signals from muons. Muons are dominant in inclined events and away from the shower core. By selecting those stations, and assuming straight line trajectories from the shower axis, the profile of muon production depth can be reconstructed~\cite{mpd} to measure $X^{\mu}_\mathrm{max}=X_1+\Delta_{\mu}$. Each hadronic interaction model predicts independent linear transformations from both $\langle X_\mathrm{max}\rangle$ and $\langle X^{\mu}_\mathrm{max}\rangle$ into an average primary mass $\langle\ln{A}\rangle$, which can then be directly compared. 

\begin{figure}[htb]
\begin{center}
\begin{minipage}[b]{0.65\linewidth}
\epsfig{file=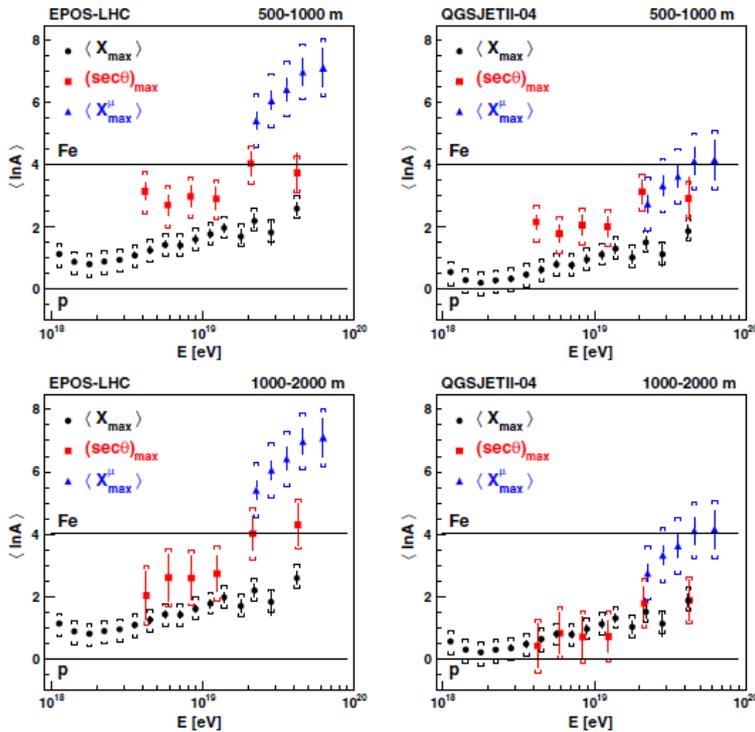,width=1.00\linewidth}
\end{minipage}
\hspace{-0.25cm}\hfill
\begin{minipage}[b]{0.35\linewidth}
\caption{Average mass from $X_\mathrm{max}$, $X^{\mu}_\mathrm{max}$ and SD asymmetry closer (top) and further (bottom) than 1~km from the shower core, measuring the internal consistency of the models EPOS-LHC (left) and QGSJetII-04 (right), from \cite{qasym}.}
\label{fig:SD}
\vspace{0.5cm}
\end{minipage}
\end{center}
\end{figure}

The same happens for the zenith angle at which the maximal asymmetry in SD signals is reached~\cite{qasym}, which is an indirect measurement of the ratio between electromagnetic and muonic signals at ground, that can be performed at different ranges of distances from the shower core. Fig. \ref{fig:SD} summarizes the results on the consistency between electromagnetic and muonic shower components: EPOS-LHC can not describe the longitudinal distribution, with a too large $\Delta_{\mu}$ for the muons surviving to the ground; while QGSJetII-04 lateral distributions lead to different ground asymmetries.

In events reconstructed by both FD and SD, the ground signals can be compared to the model predictions, given a direct electromagnetic measurement. The SD energy estimator for inclined events is obtained by scaling simulated muon maps, and the direct calibration with FD energy shows a deficit of muons in the models, which grows with energy \cite{inclined}. For other samples, the SD signals at 1000 m from the core can be compared with the corresponding predictions, by selecting showers simulated with different primaries and models that fit the measured FD profile. The expected signals are too low even when considering the heaviest mixed composition derived from FD. Analysis as a function of zenith angle shows that there is no need to rescale the electromagnetic part, but hadronic components must be increased by 30\% to 60\%~\cite{muons}.


\section{Outlook}

The hybrid detectors of the Pierre Auger Observatory have collected unprecedented amounts of data, enabling us to test the internal consistency of the hadronic interaction models used to simulate ultra high energy cosmic rays. The electromagnetic shower component has been fully characterized using FD data, and the predictions show a reasonable agreement, allowing the determination of the main features of the primary beam composition. On the other hand, the measurements done with the SD show that the hadronic shower component, and its muonic tracer, is not consistently described in any of the present models. A strong deficit of muons is seen to increase with energy and the data indicates that a rescaling of 30\% to 60\% is needed.

A combination of accelerator and cosmic ray measurements will be needed to improve the models. Auger's low energy extensions cover centre-of-mass energies similar to the LHC, in a complementary phase space region. The Observatory will take data up to 2025, with an upgraded detector, Auger Prime (Primary cosmic Ray Identification through Muons and Electrons)\cite{prime}. Scintillator detectors placed over the water Cherenkov stations will give simultaneous measurements with different sensitivity to muons, while buried scintillators will have higher muon energy thresholds. Increased FD duty-cycle (and eventually Radio measurements~\cite{radio}) may extend the hybrid data sets. 


\end{document}